\documentclass{article}

\usepackage{xcolor}
\usepackage{PRIMEarxiv}

\usepackage[utf8]{inputenc} 
\usepackage[T1]{fontenc}    
\usepackage{hyperref}       
\usepackage{url}            
\usepackage{booktabs}       
\usepackage{amsfonts}       
\usepackage{nicefrac}       
\usepackage{microtype}      
\usepackage{lipsum}
\usepackage{fancyhdr}       
\usepackage{graphicx}       
\graphicspath{{media/}}     

\title{ServerFi: A New Symbiotic Relationship Between Games and Players
}

\author{
  Pavun Shetty \\
  Yale University \\
  New Haven, CT \\
  \texttt{pavun@aya.yale.edu} \\
}

\begin{document}
\maketitle

\begin{abstract}
Blockchain-based games have introduced novel economic models that blend traditional gaming with decentralized ownership and financial incentives, leading to the rapid emergence of the GameFi sector. However, despite their innovative appeal, these games face significant challenges, particularly in terms of market stability, player retention, and the sustainability of token value. This paper explores the evolution of blockchain games and identifies key shortcomings in current tokenomics models using entropy increase theory. We propose two new models—ServerFi, which emphasizes Privatization through Asset Synthesis, and a model focused on Continuous Rewards for High-Retention Players. These models are formalized into mathematical frameworks and validated through group behavior simulation experiments. Our findings indicate that the ServerFi is particularly effective in maintaining player engagement and ensuring the long-term viability of the gaming ecosystem, offering a promising direction for future blockchain game development.
\end{abstract}


\section{Introduction}
Gaming industry ramps up as the technology keeps evolving over the whole journey packed by adventurers and outdoorsmen\cite{GAMEstanton2015brief}. Starting in the 1970s, Atari released “Pong” in the rough 1970s, an arcade table-tennis game captivated numerous consumers with  inspiring knockoff iterations.   Creations of complex, visually appealing games with richer soundscapes became possible upon more powerful microprocessors, dedicated graphics chips, and personal computers like the Commodore 64. Firmly tailing these pioneers , Nintendo took a large portion of the console market, like Duck Hunt and Excitebike, with the help of the Nintendo Entertainment System (NES) home console shortly. At the same time, Sega and Sony emerged as the competitors for their great works. Sega released the Genesis and Game Gear along with Sony’s PlayStation 2 and 3, arming CD-ROMs with enhanced game storage, together  defining what the future consoles should be like in 1994. The last one stark sign of game development appeared among that gaming chorus is the DirectX API adoption wave raised by Microsoft.

Online multiplayer games, like “World of Warcraft” and “Fortnite”, revolutionized how players interact and marked a milestone in the leap forward of the gaming industry along with the internet technology advancements that happened during real-time competition and collaboration across the globe.  With all of those became cultural phenomena, millions of players were able to share virtual worlds and fully enjoy the technology. The ruling in Google stadia and Microsoft xCloud is a lot to take in. It  streamed games directly to the players’ devices to offer high-quality gaming experiences more accessible, without the assistance of powerful hardwares \cite{GAMEgong2024impact}. These groundbreaking innovations transported the players into a world of highly social and interconnected experience, which replied on Internet technology development is undoubtedly propelling the gaming industry forward to the next era. The  visionary changes shifted public concerns to retake centralization and data ownership more seriously. During the traditional gaming time, players’ data and assets are stored onto the servers run by the game companies in an absolute centralized way, even the virtual items   were purchased by the players. The ownership of these controversial items had never been held by the ones who purchased them, under the continuous effects from   the classic economic model. It’s a model so traditional that revolves around player expenditure and company profits for decades, offering nothing to the players but little direct return upon their valuable resources investment, such as time and money. Often called “walled gardens”, having host in-game items, characters, and currencies on developers’ servers, players can not obtain ownership rights of their accounts, content, and the in-game assets. This time frame narrowed players’ rights sparing from their significant time and financial spending, and even produced no economic value to the ones that kept the financial cycle in games ongoing in stable mode and offering sustainabilities.

The appearance of GameFi resurrected the economic production relation producing real-world incentives. When it comes to combining “game” and “fi(nance)” in a more smooth way than expected, play-to-earn (P2E) games built on blockchain networks, checked all the right boxes for making a show-stopping entry.  Blockchain-based games typically cause the creation of crypto-assets in two mainstream ways: tokenizing in-game items as NFTs and granting fungible tokens the eligibility to be the currencies circulating in games \cite{PROELSS2023102916}. 
\textcolor{black}{By combining traditional gaming with on-chain assets, these games achieve decentralized ownership, transparency, and tangible economic incentives for players. However, significant challenges persist in terms of market stability, player retention, and token value sustainability.
This paper first provides an overview of the development background and seminal cases of blockchain games. We then employ entropy increase theory to analyze the underlying causes of the current challenges, elucidating the factors driving market dynamics. Building on these insights, we introduce two innovative tokenomics models: the ServerFi model of Privatization through Asset Synthesis and the Continuous Rewards for High-Retention Players model. These models are formalized into mathematical frameworks, and we validate their effectiveness through group behavior simulation experiments. Our results underscore the potential of the ServerFi model in sustaining player engagement and ensuring the long-term viability of the gaming ecosystem.}

\section{Background: Emergence of GameFi}
Blockchain-based games cause the creation of crypto-assets in two mainstream ways: tokenizing in-game items as NFTs and granting fungible tokens the eligibility to be the currencies circulating in games. 2013 had witnessed a few key moments, like Meni Rosenfeld’s concept of colored coins, awakened the emphasis on the importance of ownership of virtual assets and mirrored real-world assets onto the Bitcoin blockchain\cite{rosenfeld2012overview}. 
Larva Labs kicked off the CryptoPunks NFT collection 4 years after Meni Rosenfeld. The collection marks a significant milestone in the evolution of NFTs and inspires the ERC-721 standard for digital art and collectibles on Ethereum with no difficulty by 10,000 unique, randomly generated character images in itself\cite{NFTanselmi2023non, NFTante2023non}. 

NFT technology certainly has been sought after by a strand of visionary founders. Dapper Labs served their players in explosion with a Gamefi, the first blockchain game on Ethereum, overwhelming the network and causing significant transaction delays, named CryptoKitties. In this game, players purchase, breed, and trade virtual cats, each with unique visual characteristics of varying rarity. Their huge success highlighted the engaging nature of NFT-based gameplay. CryptoKitties capitalizes on the psychological appeal of genuine ownership and potential financial gains, attracting both avid collectors and shrewd investors within the built-in financial  cycles offering incentives in breeding and trading rare cats and creating a speculative environment. Much of the talk in the same year had been about CryptoKitties. This creative GameFi had collectively served up to millions of players engaged with the game, providing a sense of social identity and belonging across the CryptoKitties communities more than possessing these rare “kitties” merely.

While crypto games involve NFTs and P2E model, Axie Infinity from Sky Mavis, emerged as a significant successor to CryptoKitties, and it’s since become a smash hit thanks to its compelling gameplay loop that will have the players saying “just one more run” until it’s far past their bedtime. Axie Infinity allows players to collect, breed, and battle fantasy creatures well known as Axis\cite{AXIEde2022play}. There’s an NFT running behind each Axie, with unique attributes and abilities, which can be enhanced through strategic breeding and gameplay\cite{AXIEalam2022understanding}. This delightful GameFi introduces more complex game mechanics and a robust in-game economy more than merely offering economic incentives similar to CryptoKitties. Its influencing set of design ideas captivated a broad player base for this lucky pet of the age and set new bars for all the coming blockchain-based games.

\section{Tokenomics Challenges and Our Solution}

Facing plenty of competition for traditional online games running on centralized devices,
blockchain-based games are getting used to store digital assets outside the box, allowing the  items occupied by players to be sold, even transferred to another game, utilized in specific DeFis.  The incentive models are going close after massive adoption of blockchain technologies. It paved a whole new way for building cutting-edge production relationships between players and developers. Times have changed. These innovations are designed to rebuild on the electronic society that guarantees the potential to transform a booming post-gaming industry. Given these substantial advancements, one must ask: why do the game makers opt to seek a new production relationship that originated from the surging GameFi sector where players have different flavors for assets,  right after much more traditional, lean-back gaming experience in the leaping Web3 era? 

Most games have a certain lifespan, CryptoKitties is not an exception. Among the vital mechanisms work in it, the breeding, which allows players to produce new kitties, inadvertently increasing the supply and subsequently decreasing the rarity and value of individual kitties over time. Along with more players participating and breeding kitties, the secondary market becomes oversaturated shortly. The scene was novel, the players were inspired, but the dilemma was familiar enough: how to keep the price of the circulating tokens. This devaluation is further exacerbated if there are not enough active players, as the demand fails to keep pace with the growing supply.  Accordingly, individuals who allocate substantial time and resources to the process of breeding may observe a reduction in the yield of their endeavors. The initial intentional scarcity may potentially result in disenchantment and reduced levels of participation when get involved with the repeatedly appearing abundance as the game is collectively pursued.

The application of entropy increase theory alongside tokenomics provides a professional and insightful approach to elucidate the dynamics of token flow and value fluctuations within blockchain projects. Entropy increase theory, which is grounded in the second law of thermodynamics, posits that in a closed system, entropy (a measure of disorder) tends to increase over time.This concept can be analogously applied to economic systems, particularly tokenomics, to enhance our understanding of token distribution, usage, and market volatility. In tokenomics, the initial distribution of tokens is typically orderly. During this phase, tokens are relatively concentrated in the market, prices remain stable, and player expectations are high\cite{TOKENmoncadablockchain}. As time progresses, more tokens are generated through game mechanisms and enter the market. The increase in player transactions and token flow subsequently raises the market’s entropy (disorder).  During this intermediate phase, there is a surge in internal system chaos resulting in heightened volatility in token prices. Tokenomics may encounter challenges such as inflation due to an oversupply of tokens in the market and substantial price instability caused by the influx of speculators. Without effective market regulation and incentive mechanisms, the system may reach a state of high entropy (disorder), where the value of tokens generally declines, and player engagement diminishes. To maintain the long-term health of the system, it is crucial to have a way of making new incentive mechanisms and regulatory measures connect. These actions can slow the increase in entropy, thereby preserving relative market order and stability as well as maintaining player engagement.

We usually talk about tokenomics as isolated incidents, like a single point of failure with a specific cause and effect. But seen from this vantage, the story is less about any single company, and more about the all-consuming entropy of token’s circulation. Something is always devastating, some gameplays always fail out. 
Using Axie Infinity as an example, its tokenomics design presents several shortcomings from the perspective of players: Firstly, Axie Infinity’s token economy relies heavily on the continuous generation of new tokens, such as Smooth Love Potion (SLP). As more players engage and breed Axies, the number of newly generated tokens increases, leading to a rapid expansion in the market’s token supply. This supply-demand imbalance causes the value of the tokens to decline over time, resulting in the devaluation of tokens held by players.
Secondly, during the Token Generation Event (TGE), numerous players and investors flood the market with the aim of quickly profiting from buying and trading tokens. This speculative behavior can lead to significant price volatility, affecting market stability. In the long run, early speculators who profit and exit may cause token prices to plummet, adversely impacting ordinary players. Thirdly, Axie Infinity’s economic model lacks sustained incentive mechanisms to maintain player engagement post-TGE.  As the initial novelty wears off, players' enthusiasm may diminish due to limited economic incentives. Addressing any flaws in the game can help attract new users and potentially increase token demand. Participation in Axie Infinity requires players to purchase Axies, entailing a high initial investment cost. This high cost poses a barrier for new players, limiting the game’s accessibility and widespread adoption. Additionally, the market prices for rare Axies may be prohibitively high, making them unaffordable for average players.

Based on the discussion above, we have two proposals for improving the token economic model in GameFi:
\begin{enumerate}
    \item \textbf{ServerFi: Privatization through Asset Synthesis} \\
    In line with the Web3 spirit, players could be allowed to combine their in-game assets to eventually obtain sovereignty over future servers. This concept, which called “ServerFi”, involves players accumulating and merging various NFTs and other digital assets within the game to gain control over game servers. This form of privatization not only incentivizes players to invest more deeply in the game but also aligns with the decentralized and community-driven ethos of Web3. By granting players ownership and control over game servers, we can foster a more engaged and loyal player base, as they have a tangible stake in the game’s ecosystem. For example, we can design a game in which players can earn lottery chances each day based on the value they contribute to the game server. These lottery chances can be used to draw fragments. When a player collects all the necessary fragments, they can synthesize an NFT. By staking this NFT, players can share in the value contributed by users to that game server.
    
    \item \textbf{Continuous Rewards for High-Retention Players} \\
    Another approach involves the project team continuously identifying and nurturing high-retention players to maintain token vitality and ensure the game’s ecological health. By implementing sophisticated algorithms and data analytics, the project can monitor player behavior and engagement levels, providing targeted rewards and incentives to those who demonstrate strong commitment and activity within the game. This method ensures that the most dedicated players remain engaged, driving ongoing participation and interaction, which in turn supports the overall stability and growth of the game’s token economy. For example, we could design a game where a portion of the game server's revenue is airdropped daily to top-ranking users based on the value they contribute to the system. This approach would create a play-to-earn dynamic, rewarding players for their engagement and contributions.
\end{enumerate}

\section{Experiment}
To evaluate the effectiveness of our proposed tokenomics models, we conducted group behavior simulation experiments for each model. These experiments were designed to compare and analyze the differences in value capture capabilities of blockchain games built on the two distinct tokenomics frameworks. For more accurate modeling, we first formalized the definitions of these tokenomics mechanisms as follows.

\begin{enumerate}
   \item \textbf{ServerFi: Privatization through Asset Synthesis}
\begin{itemize}
    \item Let $v_i$ represent the value that player i contributes to the system in each iteration.
    \item Function $f(v) = \frac{v}{\lambda} $ denotes the number of draws a player can obtain by contributing value $v$, where $\lambda$ is a scaling constant greater than $1$.
    \item Suppose there are $k$ pieces in the lottery, and the probability of drawing each card is $1/k$.
    \item Assuming that the number of new players on the first day is $n$, and considering the game’s growth dynamics, we define the number of new players in the $i$-th iteration as $n / \alpha^{(i-1)}$.
    \item We assume that all players in the game are rational. Therefore, if a player calculates that the cost of synthesizing an NFT exceeds the current staking rewards, they will choose to leave the game. Specifically, for a new player, the expected cost to collect all fragments is $\lambda \sum_{i=1}^{k} \frac{k}{i}$. When this cost exceeds the staking rewards of a single NFT, no new users will join the game.
    \item The total value of such a system at iteration(day) $i$ is $T_i = \sum_1^n v_i$, where $n$ is the number of players at iteration $i$.
\end{itemize}
   \item \textbf{Continuous Rewards for High-Retention Players}
   \begin{itemize}
    \item Let $v_i$ represent the value that player $i$ contributes to the system in each iteration.
    \item We stipulate that the system will reward the top $20\%$ of players with $80\%$ of the total earnings from the past five days, based on their cumulative contributions during that period.
    \item We assume that all participants in the game are rational. Each player has a randomly initialized tolerance threshold, and if they fail to receive rewards multiple times consecutively, they will choose to leave the game.
    \item The total value of the system at iteration $i$ is given by $T_i = \sum_1^n v_i$, where $n$ is the number of players at iteration(day) $i$.
\end{itemize}
\end{enumerate}

\begin{figure}
  \centering
  \includegraphics[width=0.99\textwidth]{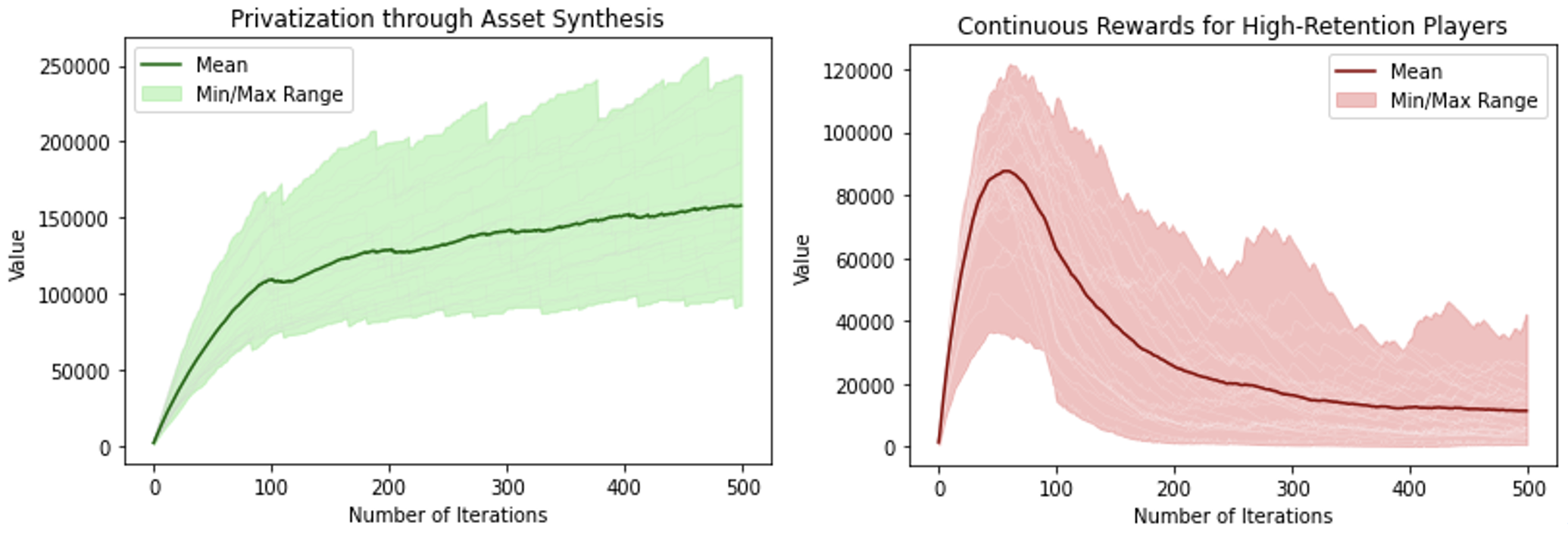}
  \caption{Player Contribution Value in Two Tokenomics Models.}
  \label{fig:fig1}
\end{figure}

Given the inherent randomness of real-world scenarios, our practical simulation experiments incorporate stochastic noise from various angles, including individual behavior and population growth. For example, we introduce mutation operators in individual modeling to capture the random fluctuations in productivity as participants engage in the game. To ensure a fair comparison between the two strategies, the experiments were designed with identical parameters, such as the maximum number of iterations and initial population size, for both experimental groups. Each economic model’s population underwent 500 iterations, with each experiment repeated 100 times. The results of the simulation are presented in Figure \ref{fig:fig1}.
The horizontal axis denotes the number of iterations, while the vertical axis represents the total value contributed by players at each iteration. The light-colored band illustrates the range between the maximum and minimum values, with the dark-colored line indicating the mean. 

In the Privatization through Asset Synthesis model(left), we observe a consistent upward trend in total player contributions as the number of iterations increases, suggesting that this model effectively sustains player engagement and drives long-term value growth. Conversely, the Continuous Rewards for High-Retention Players model(right) initially exhibits a sharp rise in player contributions, followed by a marked decline. Although this model demonstrates high player contributions in the early stages, the observed decrease over subsequent iterations indicates challenges in maintaining player engagement over the long term.

Building on the modeling results, we posit that while the strategy of continuously rewarding high-retention players may drive significant early engagement, it inherently fosters player stratification over time. Specifically, this approach risks marginalizing long-tail players by providing them with insufficient positive feedback, ultimately leading to their disengagement from the game. This stratification also tends to create high entry barriers for new players. As a result, the reduced influx of new players, coupled with the departure of long-tail players, diminishes the rewards available to existing top players, thereby perpetuating a vicious cycle.

In contrast, the ServerFi mechanism, which is grounded in fragment synthesis, introduces a degree of randomness through the fragment drawing process, thereby enhancing social mobility within the player community. For existing NFT holders, the ongoing synthesis of new NFTs ensures that even top players cannot simply “rest on their laurels”; they must consistently contribute value to maintain their status. For new or lower-contributing players, there remains a substantial opportunity to synthesize NFTs and share in server rewards, facilitating upward mobility. Thus, the ServerFi model more effectively promotes social mobility among players, revitalizing the system as a whole and fostering a more sustainable ecosystem.

\section{Conclusion}
In this paper, we have thoroughly explored the tokenomics challenges that plague current blockchain-based games. The analysis has demonstrated that traditional economic models often lead to market instability, diminishing player engagement, and unsustainable token values. To address these pressing issues, we proposed and analyzed two promising token economic models, with particular emphasis on the ServerFi model, which is based on privatization through asset synthesis. Through extensive group behavior simulation experiments, ServerFi has shown significant potential in maintaining player engagement and ensuring long-term sustainability within the gaming ecosystem. Unlike conventional models, ServerFi effectively promotes social mobility among players by introducing a dynamic and competitive environment where continuous value contribution is necessary for maintaining status. This model not only fosters a more vibrant and inclusive community but also offers a scalable and resilient framework for future blockchain games. As the industry evolves, the ServerFi approach may well represent a pivotal shift in how tokenomics is structured, providing a more sustainable path forward for the integration of decentralized technologies in gaming.





\bibliographystyle{unsrt}  
\bibliography{references}

\end{document}